\documentclass[12pt]{article}
\pdfoutput=1
\usepackage{amsmath,amsfonts,graphicx,color,bbm,tikz,bm,setspace}
\usepackage{comment}
\usepackage[nosort]{cite}
\usepackage{subfigure}
\usepackage{qcircuit}
\usepackage{multirow}
\usetikzlibrary{calc,positioning}
\usetikzlibrary{patterns,arrows,decorations.pathreplacing}
\usepackage{caption}
\usepackage{ulem}
\tikzset{>=stealth}
\usepackage{lipsum}
\usepackage{tabularx}
\usepackage{fullpage}
\usepackage{hyperref}
\usepackage{bbold}
\usepackage[english]{babel}

\newtheorem{remark}{Remark}[section]

\usepackage{listings}
\usepackage{tcolorbox}

\newcommand{\mathsym}[1]{{}}
\newcommand{\unicode}[1]{{}}

\makeatletter\@addtoreset{equation}{section}\makeatother

\newcommand{\be}{\begin{equation}}
\newcommand{\ee}{\end{equation}}
\def\beq{\begin{equation}}
\def\eeq{\end{equation}}
\newcommand{\bea}{\begin{eqnarray}}
\newcommand{\eea}{\end{eqnarray}}

\def\nn{\nonumber}

\renewcommand{\title}[1]{\vbox{\center\LARGE{#1}}\vspace{3mm}}
\renewcommand{\author}[1]{\vbox{\center{#1}}\vspace{3mm}}

\newcommand{\email}[1]{\vbox{\center\tt#1}\vspace{3mm}}

\begin{document}

\begin{center}
{\large {\bf Majorana fermions solve the tetrahedron equations\\ as well  as higher simplex equations.}}

\author{Pramod Padmanabhan,$^a$ Vladimir Korepin$^{b}$}

{$^a${\it School of Basic Sciences,\\ Indian Institute of Technology, Bhubaneswar, 752050, India}}
\vskip0.1cm
\vskip0.1cm
{$^b${\it C. N. Yang Institute for Theoretical Physics, \\ Stony Brook University, New York 11794, USA}}

\email{pramod23phys@gmail.com, vladimir.korepin@stonybrook.edu}

\vskip 0.5cm 

\end{center}


\abstract{
\noindent 
Yang-Baxter equations  define quantum integrable models. The tetrahedron and higher simplex equations are  multi-dimensional generalizations. 
Finding the solutions of these equations is a formidable task. In this work we develop a systematic method - constructing higher simplex operators
[solutions of corresponding simplex equations] from lower simplex ones. We call it lifting. By starting from solutions of Yang-Baxter equations we can construct solutions of the tetrahedron equation and simplex equation in any dimension. We then generalize this by starting from a solution of any lower simplex equation and lifting it [construct solution] to another simplex equation in higher dimension. This process introduces several constraints among the different lower simplex operators that are lifted to form the higher simplex operators. We show that braided Yang-Baxter operators 
[solutions of Yang-Baxter equations independent of spectral parameters] constructed using Majorana fermions satisfy these constraints, thus solving the higher simplex equations. As a consequence these solutions help us understand the action of an higher simplex operator on Majorana fermions. Apart from these we show that solutions constructed using Dirac (complex) fermions and Clifford algebras also satisfy these constraints. Furthermore it is observed that the Clifford solutions give rise to positive Boltzmann weights resulting in the possibility of physical statistical mechanics models in higher dimensions. Finally we also show that anti-Yang-Baxter operators [solutions of Yang-Baxter-like equations with a negative sign on the right hand side] can also be lifted to higher simplex solutions.
}

\tableofcontents 

\section{Introduction}
\label{sec:Introduction}
Since its inception in the theory of integrable quantum systems, the Yang-Baxter equation \cite{YangCN1967,BAXTER1972193,Takhtadzhan_1979,Baxter1982ExactlySM,Korepin1993QuantumIS,Franchini2016AnIT,slavnov2019algebraicbetheansatz,Retore2021IntroductionTC,Batchelor2015YangBaxterIM,Nepomechie1998ASC,eckle2019models} has found applications in many other areas, ranging from factorizable scattering theory of point particles \cite{Zamolodchikov1979FactorizedSI}, construction of knot/link polynomials \cite{Akutsu:1987bc,Akutsu:1987ea,Akutsu:1987qs,Turaev1988TheYE,Kauffman1991}, AdS/CFT integrability \cite{Beisert2010ReviewOA,deLeeuw2019OnepointFI}, quantum information and computation \cite{Aravind1997,Padmanabhan_2020,Padmanabhan_2021,Padmanabhan2019QuantumES,Quinta_2018,Rowell2007ExtraspecialTG,Zhang_2024,zhang2005yang,kauffman2004braiding,zhang2013integrable,kauffman2002quantum,kauffman2003entanglement}, and in quantum circuits to describe integrable and open quantum systems \cite{Vanicat2017IntegrableTL,VanDyke2021PreparingBA,aleiner2021bethe,Claeys:2021cdy,sa2021integrable,PhysRevLett.126.240403,10.21468/SciPostPhys.8.3.044,su2022integrable,Miao2022TheFB,Gombor2022IntegrableDO,Sopena_2022,Miao2023IntegrableQC,zhang2024geometricrepresentationsbraidyangbaxter,znidaric2024integrabilitygenerichomogeneousu1invariant}. Yang-Baxter operators, which are the solutions of the Yang-Baxter equations, come with two indices. For the representations useful in physics, Hilbert spaces are attached to sites with these indices. These operators can be extended to include more indices and made to satisfy corresponding constraint equations that generalize the Yang-Baxter equation. The first instance of such an equation is known as the tetrahedron equation \cite{Zamolodchikov1981TetrahedronEA,Zamolodchikov1980TetrahedronOriginal}. In its original formulation it appears as a constraint on the factorizable scattering of straight lines. Taking cues from the Yang-Baxter case, the tetrahedron equation was subsequently pursued to study three dimensional integrable models \cite{Bazhanov1982ConditionsOC,Bazhanov1984FreeFO,Maillet1989IntegrabilityFM,Bazhanov1992NewSL,Bazhanov1993StartriangleRF,Sergeev1995TheVF,Stroganov1997TetrahedronEA,Mangazeev2013AnI3,Khachatryan2015IntegrabilityIT,Talalaev2015ZamolodchikovTE,talalaev2018integrablestructure3dising,Kuniba2015MultispeciesTA,khachatryan2023integrability,BOOS_1995}. Following this the focus has mostly been on solving the tetrahedron equation \cite{Baxter1983OnZS,MAILLET1989221,Frenkel1991SimplexEA,Hietarinta_1993,CarterSaito,Hietarinta_1997,WadatiShiroishi1995,Liguori_1993,Gavrylenko2020SolutionOT,bardakov2022settheoretical,Kuniba2022NewST,sun2022cluster,inoue2024solutions,inoue2024tetrahedron,iwao2024tetrahedronequationschurfunctions} and studying the algebraic structures associated with them \cite{Kapranov19942CategoriesAZ,Baez1995HigherDA,Bazhanov2005ZamolodchikovsTE,Bazhanov2009QuantumGO,Kuniba2015TetrahedronEA,Talalaev2021TetrahedronEA,kuniba2022quantum}. However there was a significant limitation to many of the solutions as they did not yield positive Boltzmann weights and hence were deemed unphysical. We will see that some of the solutions obtained here do result in positive Boltzmann weights making them potential candidates for higher dimensional integrable statistical mechanics models. 

It is desirable to extend the range of applications of the higher simplex equations to mimic the success in the Yang-Baxter case. This requires systematic methods to construct solutions that can be applicable in quantum computing and quantum information and not just in higher dimensional integrable models. Moreover we want to develop methods that are not restricted just to the tetrahedron equation but that can also solve the simplex equations in any dimension. 

We propose such a method in this work, where we use solutions of lower simplex equations to construct solutions of higher simplex equations. This means that we use a $k$-simplex solution and all the other simplex operators below this, to construct solutions of a $d$-simplex equation, for $d>k$. For instance we can use a 1- and a 2-simplex operator [solutions of the 1-simplex and 2-simplex (Yang-Baxter) equations] to construct solutions of the tetrahedron and other higher simplex equations. We call this procedure the {\it lifting} method. In this process several constraints on the lower simplex operators naturally arise due to the index structure of the $d$-simplex equation. These constraint equations resemble modified simplex equations. The challenge then becomes finding lower simplex operators that satisfy these constraints. We overcome this by finding different solutions of these constraints using physically relevant algebras, like those of Majorana and Dirac fermions and from other Clifford algebras, thus making the lifting procedure successful. 

Apart from usual lower simplex solutions we find that it is possible to lift {\it anti-lower simplex solutions} to higher simplex solutions broadening the scope of the lifting method. The latter solve the anti-lower simplex equations introduced in \cite{padmanabhan2024solving}. These are just the usual lower simplex equations but with the right hand side multiplied by an overall negative sign. These equations can be generalized to include other roots of unity in place of the negative sign, but we do not consider those here.

The contents are laid out as follows. We begin by analyzing the lifting of the 1-simplex and the Yang-Baxter (2-simplex) operators. We state the different ans\"{a}tze used in constructing the higher simplex solutions and the constraint equations that arise as a consequence of demanding these ans\"{a}tze to be solutions in Sec. \ref{sec:constraints}. For the simplest case of lifting Yang-Baxter solutions we find just two constraints. In Sec. \ref{sec:solutions} we discuss the solutions that satisfy these constraints. Some of the solutions obtained by lifting Yang-Baxter operators yield higher simplex solutions with positive Boltzmann weights. This is discussed in Sec. \ref{sec:positivity}. Following this we proceed to lifting tetrahedron operators to higher simplex solutions in Sec. \ref{sec:tetrahedron}. We encounter three more constraints in addition to the two before. The solutions for these constraints are also provided here, showing that we can successfully lift the tetrahedron operator in different ways. The ideas from both these cases are then generalized to the lifting of a simplex solution in arbitrary dimension to a higher dimension solution in Sec. \ref{sec:generalLifts}. We count and write down the general set of constraint equations that arise in this case. In Sec. \ref{sec:lifting-antiYBO} we show ways in which the anti-Yang-Baxter operator can be lifted into higher simplex solutions. We end with a summary and scope for future work in Sec. \ref{sec:conclusion}. We also briefly analyze the `braiding' action of the tetrahedron operators on the Majorana fermions here. 

\section{Constraints}
\label{sec:constraints}
We will now derive the constraints on the lower simplex operators (Yang-Baxter or 2-simplex operators and 1-simplex operators), that need to be satisfied to obtain the corresponding solutions to the higher simplex equations.  We work with the constant vertex form of the higher simplex equations \cite{MAILLET1989221,Hietarinta_1994,Hietarinta1997TheET} that describes the scattering of the $\frac{d(d+1)}{2}$ particles formed at the vertices of the $d+1$ strings in $d$ dimensions, when spectral parameters are introduced. But first we fix the notation to be followed in the rest of the paper. 

\paragraph{Notation :} The operators used in this paper come with multiple indices, with each of them indexing a local Hilbert space denoted $V$. The simplest qubit or two dimensional representation corresponds to the choice $V\simeq\mathbb{C}^2$. The values for these indices range from $\{1,2,\cdots, N\}$, similar to what appears in the spin chain and braid group literature. Keeping this in mind we will also call these local Hilbert spaces as sites. In this setup we have the following set of operators -
\begin{eqnarray}
    M_j~~;~~Y_{jk},
\end{eqnarray}
acting on 1-, and 2-sites respectively. They satisfy the constant forms of the 1- and 2-simplex (Yang-Baxter) equations, 
\begin{equation}\label{eq:1-simplex}
    M_iN_i = N_iM_i,~~~\text{or}~~~M_i^2=M_i^2
\end{equation}
and 
\begin{equation}\label{eq:2-simplex}
    Y_{ij}Y_{ik}Y_{jk} = Y_{jk}Y_{ik}Y_{ij},
\end{equation}
respectively. Here $N_j$ is another single site operator.
Other higher or $d$-simplex operators starting from the tetrahedron or 3-simplex operator will be denoted as $$ R_{i_1i_2\cdots i_d}.$$

We will now discuss the possible ways of constructing a given $d$-simplex operator from the 1- and 2-simplex operators, $M$ and $Y$. This will be shown for the $d=2,3,4$-simplex equations. In the process we arrive at the constraints that $M$ and $Y$ have to satisfy in order to successfully become a higher simplex operator. The index structure of the higher simplex equations ensure that these constraints remain unchanged for the $d>4$ case as well.

\subsection{Yang-Baxter or 2-simplex equations}
\label{subsec:constraints-2-simplex}
The Yang-Baxter or 2-simplex operator can be built out of 1-simplex operators as follows
\begin{equation}\label{eq:11to2-simplexOP}
    Y_{jk} = M_jM_k.
\end{equation}
This trivially satisfies the constant 2-simplex equation when substituted into its vertex form \eqref{eq:2-simplex}. Thus in this case we do not require any constraint on the 1-simplex operators $M_j$ to obtain a solution of the 2-simplex equation. The assumption $\left[M_i, M_j\right]=0$ for $i\neq j$, is made in arriving at this conclusion. However this is not the case when the $M_j$'s are realized using Majorana fermions, the main focus of this work, as then we will have $\{M_i, M_j\} =0$. We will return to this point in Sec. \ref{sec:solutions}.

A natural generalization of the solution in \eqref{eq:11to2-simplexOP}
\begin{equation}\label{eq:11to2-simplexOP-2}
    Y_{jk} = M_jN_k,
\end{equation}
subject to the constraint 
\begin{equation}
    \left[M, N\right]=0,
\end{equation}
becomes a 2-simplex operator. Such solutions were also seen in \cite{padmanabhan2024solving}. 

Finally we note that these are the only possibilities for constructing a 2-simplex operator from lower simplex ones, namely the 1-simplex operators. This can also be seen as a partition of the integer 2. 

\subsection{Tetrahedron or 3-simplex equations}
\label{subsec:constraints-3-simplex}
There are two ways to partition 3 into 2 and 1. We have 
\begin{eqnarray}
    R_{ijk} & = & M_iM_jM_k \label{eq:111to3simplex} \\
    R_{ijk} & = & M_iY_{jk}~~;~~Y_{ij}M_k. \label{eq:21to3simplex}
\end{eqnarray}
We require them to solve the vertex form of the constant 3-simplex or tetrahedron equation 
\begin{equation}\label{eq:3-simplex}
    R_{ijk}R_{ilm}R_{jln}R_{kmn} = R_{kmn}R_{jln}R_{ilm}R_{ijk}.
\end{equation}
While the solutions \eqref{eq:111to3simplex} trivially solve the 3-simplex equation in \eqref{eq:3-simplex}, the operators in \eqref{eq:21to3simplex}\footnote{Versions of the solution \eqref{eq:21to3simplex} can also be found in \cite{Hietarinta_1993} and further developed in \cite{singh2024unitarytetrahedronquantumgates} for the constant $4\times 4 $ Yang-Baxter solutions of Hietarinta \cite{Hietarinta_1992}.} solve the 3-simplex equation when 
\begin{equation}\label{eq:YMMconstraint-1}
    M_iM_jY_{ij} = Y_{ij}M_jM_i.
\end{equation}
Henceforth we denote this the {\it YM-constraint}.
When $\left[M_i, M_j\right]=0$ this condition reduces to 
\begin{equation}\label{eq:YMMconstraint-2}
\left[Y_{ij}, M_iM_j \right]=0~~\textrm{or}~~\left[Y, M\otimes M\right] = 0.
\end{equation}
As observed in the Yang-Baxter case (See Sec. \ref{subsec:constraints-2-simplex}), when $\{M_i, M_j\}=0$ as in the case of Majorana fermions, the condition \eqref{eq:YMMconstraint-1} should be satisfied in place of the condition \eqref{eq:YMMconstraint-2}. We work with the former as it includes both cases.

The logic behind the choices in \eqref{eq:21to3simplex} can be understood with a closer look on the index structure of the 3-simplex equation \eqref{eq:3-simplex}. We see that the sets of indices $\{ij, il, jl\}$ and $\{lm, ln, mn\}$ are the same as those appearing in the vertex form of the 2-simplex equation \eqref{eq:2-simplex} or the Yang-Baxter equation in the non-braided form. Thus it is natural to expect the operators in \eqref{eq:21to3simplex} to satisfy the 3-simplex equation when the YM-constraint is satisfied.

Other possibilities for 3-simplex operators are considered in the following remarks.
\begin{remark}
 The operator \eqref{eq:111to3simplex} can be generalized to 
$$ M_iN_jQ_k.$$
This naturally introduces constraints among the three operators $M$, $N$ and $Q$,
\begin{equation}
    N_jM_jQ_kM_kQ_nN_n = M_jN_jM_kQ_kN_nQ_n.
\end{equation}
There are multiple ways to satisfy this constraint, however we will not focus on such solutions in this work, sticking to the simplest solutions in \eqref{eq:111to3simplex}.   
\end{remark}

\begin{remark}
Taking into account the index structure of the 3-simplex equation \eqref{eq:3-simplex}, we could also consider ans\"{a}tze built from two Yang-Baxter operators
\begin{equation}
        Y_{ij}Y_{jk},
\end{equation}
or more generally 
\begin{equation}
        Y_{ij}\Tilde{Y}_{jk}.
\end{equation}
The former satisfies the 3-simplex equation when 
\begin{equation}
    Y_{il}Y_{ln} = Y_{ln}Y_{il},
\end{equation}
while the latter needs to satisfy additional constraints akin to the one in \eqref{eq:YYconstraint} to result in a 3-simplex solution.
\end{remark}

\begin{remark}\label{rem:forbiddenYM}
    The final possibility involves substituting the operator $$ Y_{ik}M_j$$ into the 3-simplex equation \eqref{eq:3-simplex}. However in this case the pair of indices on $Y$ is included in the set $\{ik, im, jn, kn\}$ appearing in the 3-simplex equation. As is evident, this set does not contain the index structure of the 2-simplex equation. This implies that the operator $Y$ is not necessarily a 2-simplex operator and consequently the operator $Y_{ik}M_j$ has to satisfy more complicated constraints to become a 3-simplex solution. Such possibilities, where the indices on $Y$ are not `nearest-neighbors' in the higher simplex equation, occur for all $d$-simplex equations. As these index pairs belong to a set of indices that do not coincide with the index structure of the 2-simplex equation, we omit these possibilities in the rest of this work.
\end{remark}

\subsection{4-simplex equations}
\label{subsec:constraints-4-simplex}
The number of partitions increase with $d$ as expected. For the 4-simplex operator we have
\begin{eqnarray}
    R_{ijkl} & = & M_iM_jM_kM_l \label{eq:1111to4-simplex} \\ 
    R_{ijkl} & = & Y_{ij}M_kM_l~~;~~M_iY_{jk}M_l~~;~~M_iM_jY_{kl} \label{eq:211to4-simplex} \\
    R_{ijkl} & = &  Y_{ij}Y_{kl} \label{eq:22to4simplex}.
\end{eqnarray}
As in the previous two cases the solutions in \eqref{eq:1111to4-simplex} constructed out of the 1-simplex operators $M$, trivially solve the vertex form of the 4-simplex equation
\begin{eqnarray}\label{eq:4-simplex}
    R_{ijkl}R_{imnp}R_{jmqr}R_{knqs}R_{lprs} = R_{lprs}R_{knqs}R_{jmqr}R_{imnp}R_{ijkl}.
\end{eqnarray}
The operators in \eqref{eq:211to4-simplex}, built using the $Y$ and $M$ operators, satisfy the 4-simplex equation when the YM-constraint is fulfilled. The rationale for expecting such solutions is similar to the tetrahedron case and can be deduced from the index structure of the 4-simplex equation \eqref{eq:4-simplex}. More precisely the index sets $\{ij, im, jm\}$, $\{mn, mq, nq\}$ and $\{qr, qs, rs\}$ coincide with the index structure of the 2-simplex equations. This justifies the choice of operators in \eqref{eq:211to4-simplex}, \eqref{eq:22to4simplex}.

The most interesting case is the third choice constructed out of two Yang-Baxter operators \eqref{eq:22to4simplex}. Plugging in this ansatz into the 4-simplex equation yields a new constraint among the two Yang-Baxter operators
\begin{equation}\label{eq:YYconstraint}
   Y_{ij}Y_{kl}Y_{ik}Y_{jl} = Y_{jl}Y_{ik}Y_{kl}Y_{ij}. 
\end{equation}
This will be denoted the {\it YY-constraint}.
We will soon see that the YM- and the YY-constraints are the only ones that need to be satisfied for the construction of $d$-simplex operators when $d>4$. This is again a consequence of the index structure of these equations. A couple of remarks are in order.
\begin{remark}
    The operator ans\"{a}tze in \eqref{eq:1111to4-simplex}, \eqref{eq:211to4-simplex} and \eqref{eq:22to4simplex} can be generalized to 
    \begin{eqnarray}
    R_{ijkl} & = & M_iN_jQ_kS_l \nonumber \\ 
    R_{ijkl} & = & Y_{ij}M_kN_l~~;~~M_iY_{jk}N_l~~;~~M_iN_jY_{kl} \nonumber \\
    R_{ijkl} & = &  Y_{ij}\Tilde{Y}_{kl} \nonumber.
\end{eqnarray}
They need to satisfy more constraints, similar to the ones seen for the 3-simplex equation, to turn into a genuine 4-simplex operator. We will not consider them here other than just noting these possibilities.
\end{remark}

\begin{remark}\label{rem:P-YY}
    We find that the permutation generators $P_i\equiv P_{i,i+1}$ satisfying the relations
    \begin{eqnarray}\label{eq:permRelations}
        P_iP_{i+1}P_i & = & P_{i+1}P_iP_{i+1}, \nonumber \\
        P_i^2 & = & \mathbb{1}, \nonumber \\
        P_iP_j & = & P_jP_i,~~|i-j|>1,
    \end{eqnarray}
    satisfy the YY-constraint upon the identification
    $$ Y_{ij} = P_{ij}.$$ This property becomes crucial in Sec. \ref{sec:solutions} when we consider the Majorana solutions.
    
    Moreover they also satisfy the YM-constraint for any $M$, which just follows from the definition of the permutation operator. This implies that the generators of the permutation group can be lifted to higher simplex operators for an arbitrary choice of $M$.
    
\end{remark}

\subsection{$d$-simplex equations}
\label{subsec:constraints-d-simplex}
Now we partition $d$ into 2's and 1's. Thus a $d$-simplex operator is made from $s$, $Y$'s and $(d-2s)$ $M$'s, with $s\in\{0,1,\cdots, \frac{d}{2}\}$ and $s\in\{0,1,\cdots, \frac{d-1}{2}\}$, when $d$ is even and odd respectively.

The $d$-simplex operators can then be constructed using these $Y$'s and $M$'s after ensuring that the index pair on the $Y$'s are such that they are `nearest-neighbors' on the $d$-simplex operator. The notion of `nearest-neighbor' indices is easily seen through examples. For instance consider a 5-simplex operator $R_{ijklm}$. When $s=1$, the allowed index pairs on the lone $Y$ are $(ij)$, $(jk)$, $(kl)$ and $(lm)$. And when $s=2$, the allowed indices on the two $Y$'s are $(ij ; kl)$, $(ij ; lm)$ and $(jk ; lm)$. Some of the forbidden pairs when $s=1$ are $(ik)$, $(il)$ or $(jm)$ and when $s=2$ some of the forbidden ones are $(ij ; km)$, $(im ; jl)$ etc. The reason for avoiding these index pairs is discussed in Remark \ref{rem:forbiddenYM} in the context of 3-simplex equations. Furthermore the above prescriptions help us to find the total number of $d$-simplex operators for a given $s$ or the number of $Y$'s. For instance when $s=1$ there are $d-1$ $d$-simplex operators and when $s=2$ there are $\frac{(d-3)(d-2)}{2}$ $d$-simplex operators.

With these configurations of the $Y$ and $M$ operators we can construct the $d$-simplex solutions when the YM- and YY-constraints are satisfied. These are repeated here for convenience :
\begin{eqnarray}\label{eq:YY-YM-constraints}
    M_iM_jY_{ij} & = & Y_{ij}M_jM_i, \nonumber \\
    Y_{ij}Y_{kl}Y_{ik}Y_{jl} & = & Y_{jl}Y_{ik}Y_{kl}Y_{ij}.
\end{eqnarray}
Choosing forbidden index pairs for the $Y$ operators introduces more complicated constraints that do not resemble the 2-simplex equation and so we avoid such ans\"{a}tze. 

Before proceeding further we show that the symmetries of the Yang-Baxter or 2-simplex equation carries over to the YM- and YY-constraints. The Yang-Baxter equation has two types of symmetries : continuous and discrete symmetries \cite{hietarinta1993-JMP-Long,Hietarinta1993-BookChapter,maity2024algebraic}. The continuous symmetries is generated by a set of invertible operators $Q$ and a constant $\kappa$'s as
\begin{eqnarray}\label{eq:contSymmetries}
    & Y \to \kappa_1~(Q\otimes Q)Y(Q\otimes Q)^{-1}\,, & \nonumber \\
    & M \to \kappa_2~Q~M~Q^{-1}\, . &
\end{eqnarray}
A small computation shows that this leaves both the YM- and YY-constraints invariant. As in the Yang-Baxter case we call these local or gauge symmetries of these system of equations. Note that the vertex forms of the higher simplex equations also possess these symmetries with the appropriate modifications.

There are discrete symmetries of the Yang-Baxter equations. The first corresponds to transposing the $Y$ matrix. For the constraints, this corresponds to transposing both the $Y$ and $M$ matrices 
\begin{eqnarray}\label{eq:Discrete-I}
    Y\rightarrow Y^T~~;~~M\rightarrow M^T,~~~\textrm{Discrete-I}.
\end{eqnarray}
The second discrete symmetry corresponds to a relabelling the basis of the local Hilbert space, or a permutation of the indices of the basis elements. When the local Hilbert space is $\mathbb{C}^2$ this symmetry can be written as 
\begin{eqnarray}\label{eq:Discrete-II}
    Y \rightarrow \left(X\otimes X\right)Y\left(X\otimes X\right),~~~\textrm{Discrete-II}.
\end{eqnarray}
Here $X$ is the first Pauli matrix, also thought of as the shift operator of $\mathbb{Z}_2$. Clearly this is a special case of the continuous symmetries \eqref{eq:contSymmetries}. Adding $M\rightarrow XMX$ to this we obtain the symmetries of the two constraint equations. This symmetry can be extended to an arbitrary qudit space $\mathbb{C}^d$ by replacing the $X$ operator with the appropriate $\mathbb{Z}_d$ shift operator.

The third discrete symmetry of the Yang-Baxter equation corresponds to interchanging the spaces on which the 2-simplex operator $Y$ acts. That is
\begin{eqnarray}\label{eq:Discrete-III}
    Y \rightarrow PYP,~~~\textrm{Discrete-III}
\end{eqnarray}
where $P$ is the permutation operator satisfying the relations \eqref{eq:permRelations}. This is also a symmetry of the YY-constraint as this transformation results in the equation
\begin{eqnarray}
     Y_{ji}Y_{lk}Y_{ki}Y_{lj}  =  Y_{lj}Y_{ki}Y_{lk}Y_{ji},
\end{eqnarray}
which is obtained from the original equation with the interchanges $i\leftrightarrow l$ and $j\leftrightarrow k$.
Next the YM-constraint is left invariant when $M_i\leftrightarrow M_j$ along with the transformed $Y$. Thus we see that all the symmetries of the original Yang-Baxter equation carry over to the constraint equations.

Thus with just two constraints for the $Y$'s and $M$'s we can construct the solutions of the vertex form of all the $d$-simplex equations. Our next task is to find the $Y$'s and $M$'s which satisfy these constraints. 

\section{Solutions}
\label{sec:solutions}
All of the solutions that we write down are algebraic, in the sense that they are independent of the choice of the dimension of the local Hilbert space $V$ (representation independence). With this in mind we will use the following algebras to construct the $Y$'s and $M$'s :
\begin{enumerate}
    \item The algebra of Majorana fermions.
    \item The algebra of Dirac [complex] fermions. 
\end{enumerate}
Apart from these we will also identify the Yang-Baxter operators among the Hietarinta's classification \cite{Hietarinta_1992}, that satisfy the YY-constraint. Following this the corresponding $M$ operators satisfying the YM-constraint are found. Here too the algebraic version will be considered using the expressions from \cite{maity2024algebraic}.

Note that the solutions obtained from the Majorana and Dirac fermions have well-defined actions on the occupation number basis or the Fock space. In these cases the indices appearing in the simplex equations index the fermions.

\subsection{From Majorana fermions}
\label{subsec:majoranafermion}
Consider a set of $N$ Majorana fermions $\{\gamma_j|j\in\{1,\cdots, N\} \}$ satisfying the relations
\begin{eqnarray}\label{eq:majoranaRelations}
    \gamma_j = \gamma_j^\dag~~;~~\{\gamma_j, \gamma_k\}=2\delta_{jk},~~j,k\in\{1,\cdots, N\}.
\end{eqnarray}
These operators can be thought of as real fermions and also as generators of a Clifford algebra.

The following operator constructed using these Majorana fermions
\begin{eqnarray}\label{eq:majoranaBraid}
    \sigma_{jk} = \mathbb{1}+\gamma_j\gamma_k,
\end{eqnarray}
satisfies the braid relation
\begin{eqnarray}
    \sigma_{i}\sigma_{i+1}\sigma_{i} = \sigma_{i+1}\sigma_{i}\sigma_{i+1},~~\sigma_i\equiv\sigma_{i,i+1}.
\end{eqnarray}
This solution first appeared in the context of non-Abelian statistics of $p$-wave superconductors \cite{Ivanov_2001} and in fermionic quantum computation \cite{Bravyi_2002}. Subsequently it was used as a quantum gate and Baxterized in \cite{kauffman2013knotlogictopologicalquantum,Kauffman_2016,Kauffman_majorana}. The inverse of this operator is given by 
\begin{eqnarray}
    \sigma_{jk}^{-1} = \frac{1}{2}\left(\mathbb{1} - \gamma_j\gamma_k\right),
\end{eqnarray}
and $\sigma^8\propto\mathbb{1}$.

We will now verify that the 2-simplex operator $Y$, built from $\sigma$  satisfies the YY-constraint. To this end the $Y$ operator takes the form 
\begin{eqnarray}\label{eq:Ymajorana}
    Y_{jk} = P_{jk}\sigma_{jk} = P_{jk}~\left[\mathbb{1}+\gamma_j\gamma_k \right],
\end{eqnarray}
where the $P$'s are permutation operators satisfying the relations \eqref{eq:permRelations}. Substituting this into the YY-constraint we find that the left hand side simplifies to 
$$ 2P_{ij}P_{kl}P_{ik}P_{jl}~\left[\gamma_i\gamma_j + \gamma_i\gamma_k + \gamma_j\gamma_l + \gamma_k\gamma_l \right],$$
and the right hand simplifies to 
$$ 2P_{jl}P_{ik}P_{kl}P_{ij}~\left[\gamma_i\gamma_j + \gamma_i\gamma_k + \gamma_j\gamma_l + \gamma_k\gamma_l \right].$$
However by Remark \ref{rem:P-YY} these two are equal and hence the 2-simplex operator in \eqref{eq:Ymajorana} satisfies the YY-constraint. 

Having ascertained the $Y$ operator \eqref{eq:Ymajorana}, we can proceed to compute the corresponding $M$ operators such that they satisfy the YM-constraint. This is simply achieved by the choice
\begin{eqnarray}\label{eq:Mmajorana}
    M_j = \gamma_j.
\end{eqnarray}
It is easy to verify that both sides of the YM-constraint simplify to 
$$P_{ij}~\left[\mathbb{1} - \gamma_i\gamma_j\right]. $$

To summarize we find that the operators 
$$ Y_{jk} = P_{jk}~\left[\mathbb{1}+\gamma_j\gamma_k \right]~~;~~M_j=\gamma_j$$
satisfy both the YM- and YY-constraints and hence can be used to construct $d$-simplex operators as considered in Sec. \ref{subsec:constraints-d-simplex}.

\subsection{From Dirac [complex] fermions}
\label{subsec:complexfermion}
For the next set of solutions, we consider the algebra of $N$ fermions satisfying the relations
\begin{eqnarray}
   & \{a_j, a_k^\dag\} = \delta_{jk} & \nonumber \\
   & \{a_j, a_k\} = 0~~;~~\{a_j^\dag, a_k^\dag\} = 0,~~~j,k\in\{1,\cdots, N \}. \label{eq:fermionRelations} &
\end{eqnarray}
In this setting we can construct two $Y$ operators that satisfy the vertex form of the 2-simplex equation \eqref{eq:2-simplex} or the non-braided Yang-Baxter equation
\begin{eqnarray}
    Y_{jk} & = & \mathbb{1} + \alpha~a_ja_k, \label{eq:Yfermion-1} \\
     Y_{jk} & = & \mathbb{1} + \alpha~P_{jk}a_ja_k, \label{eq:Yfermion-2}
\end{eqnarray}
with $\alpha\in\mathbb{C}$ and $P$ the permutation operator. It is simple to see that two more $Y$ operators can be constructed by replacing $a$ with $a^\dag$ in the above expressions. Note also that both these operators are invertible and they are obtained by replacing $\alpha$ with $-\alpha$ in the corresponding expressions.

Both these operators satisfy the YY-constraint. This can be easily verified that for the $Y$ operator in \eqref{eq:Yfermion-1}, as sides of the constraint equation simplify to
$$ \mathbb{1} + \alpha~\left[a_ia_j + a_ka_l + a_ia_k + a_ja_l \right],$$
and for the $Y$ operator in \eqref{eq:Yfermion-2}, the constraint equation reduces to 
$$\mathbb{1} + \alpha~\left[P_{ij}a_ia_j + P_{kl}a_ka_l + P_{ik}a_ia_k + P_{jl}a_ja_l \right] + \alpha^2~\left[P_{ij}P_{kl}a_ka_la_ia_j + P_{jl}P_{ik}a_ja_la_ia_k\right].$$

We can now find the $M$ operators corresponding to these $Y$ operators such that the YM-constraint is satisfied. We find that in both cases the operator
\begin{eqnarray}\label{eq:Mfermion}
    M_j = \mathbb{1} - 2a_j^\dag a_j,
\end{eqnarray}
is a valid choice. This operator squares to the identity making it invertible. Thus the $Y$ operators in \eqref{eq:Yfermion-1} and \eqref{eq:Yfermion-2} and the $M$ operators in \eqref{eq:Mfermion} help construct higher simplex operators using the prescriptions of Sec. \ref{sec:constraints}. The following remarks are important :

\begin{remark}
    The adjoint of \eqref{eq:Yfermion-1} and \eqref{eq:Yfermion-2} are also solutions. This can be seen as an extension of the first discrete symmetry \eqref{eq:Discrete-I} to include complex conjugation along with the transpose. Thus the solutions in \eqref{eq:Yfermion-1} and \eqref{eq:Yfermion-2} and their adjoints are equivalent.
\end{remark}

\begin{remark}
    It is well known that Majorana fermions can be constructed out of complex fermions
    \begin{eqnarray}\label{eq:d2m}
        \gamma_j = a_j + a_j^\dag~~;~~\tilde{\gamma}_j = \mathrm{i}~\left(a_j - a_j^\dag\right).
    \end{eqnarray}
    Thus for every complex fermion we can construct two Majorana fermions. Substituting these into the fermionic $Y$ and $M$ constructed above helps us view these solutions as being constructed out of Majorana fermions, $\gamma$ and $\tilde{\gamma}$.
\end{remark}

\begin{remark}
   We could also choose $M_j=a_j$ along the lines of Majorana case. However they do not satisfy the YM-constraint. Nevertheless the resulting higher simplex operators still solve the appropriate equations due to the appearance of $M^2$ on both sides of these equations. While these are valid higher simplex solutions they are non-invertible.  
\end{remark}

\begin{remark}
    With the introduction of spectral parameters into the higher simplex equations it is possible to obtain more solutions. One such example is given by 
    \begin{eqnarray}
        Y_{jk}(u) = P_{jk} + \alpha u~a_ja_k,~~\alpha, u\in\mathbb{C}.
    \end{eqnarray}
    This operator squares to the identity operator and satisfies the spectral parameter dependent 2-simplex equation
    \begin{eqnarray*}
        Y_{ij}(u)Y_{ik}(u+v)Y_{jk}(v) = Y_{jk}(v)Y_{ik}(u+v)Y_{ij}(u).
    \end{eqnarray*}
    This operator can also be lifted to higher simplex solutions provided they satisfy the spectral parameter dependent form of the YM- and YY-constraints :
    \begin{eqnarray}
        M_jM_kY_{jk}(z) & = & Y_{jk}(z)M_kM_j, \nonumber \\
        Y_{ij}(z)Y_{kl}(z)Y_{ik}(\tilde{z})Y_{jl}(\tilde{z}) & = & Y_{jl}(\tilde{z})Y_{ik}(\tilde{z})Y_{kl}(z)Y_{ij}(z),
    \end{eqnarray}
    where $z$, $\tilde{z}$ are complex spectral parameters and $M$ is independent of the spectral parameter. The spectral parameter dependent $Y$ operator satisfies the second constraint. For this choice of $Y$, the possible $M$ operators are 
    \begin{eqnarray}
        M_j = \mathbb{1} - 2a_j^\dag a_j~~;~~M_j = a_j,
    \end{eqnarray}
    giving invertible and non-invertible higher simplex solutions respectively. The resulting higher simplex operators satisfy spectral parameter dependent higher simplex equations. While it is cumbersome to write these modified equations for an arbitrary $d$, we illustrate the dependencies by writing down the 3- and 4-simplex equations with spectral parameters
    \begin{eqnarray}
       & R_{ijk}(u)R_{ilm}(u+v)R_{jln}(v)R_{kmn}(z) = R_{kmn}(z)R_{jln}(v)R_{ilm}(u+v)R_{ijk}(u) & \nonumber \\
       &  R_{ijkl}(u,z)R_{imnp}(u+v,z)R_{jmqr}(v,\tilde{u})R_{knqs}(\tilde{z}, \tilde{u}+\tilde{v})R_{lprs}(\tilde{z}, \tilde{v}) = & \nonumber \\ 
       &R_{lprs}(\tilde{z}, \tilde{v})R_{knqs}(\tilde{z}, \tilde{u}+\tilde{v})R_{jmqr}(v,\tilde{u})R_{imnp}(u+v,z)R_{ijkl}(u,z). &
    \end{eqnarray}
\end{remark}

\subsection{From Hietarinta's constant Yang-Baxter solutions}
\label{subsec:hietarinta}
In the early 90's\footnote{There were several attempts to classify two dimensional Yang-Baxter solutions preceding the work of Hietarinta \cite{Sogo,Hlavaty_1987,Fei:1991zn,Zhang_1991}.} Jarmo Hietarinta, in a series of papers \cite{Hietarinta_1992,hietarinta-conferenceProceeding,hietarinta1993-JMP-Long,Hietarinta1993-BookChapter}, classified the $4\times 4$ or two dimensional solutions of the constant [independent of spectral parameter] Yang-Baxter equation. The solutions cover operators of all possible ranks, including both invertible and non-invertible solutions. We now ask the question which of the full rank solutions can be lifted to solve the higher simplex equations. In the current context this translates to identifying the invertible solutions that can satisfy the YY-constraint. Following this we find the corresponding $M$ operators by imposing the YM-constraint.  

A constant solution of the 2-simplex or Yang-Baxter equation implies an equivalence class of operators generated by the continuous \ref{eq:contSymmetries} and discrete symmetries \eqref{eq:Discrete-I}-\eqref{eq:Discrete-III} of the equation. Hietarinta finds 10 such classes for the invertible case. A representative element of each of these classes is shown below
\begin{eqnarray}
    & & H3,1 = \begin{pmatrix} k & 0 & 0 & 0 \\ 0 & q & 0 & 0 \\ 0 & 0 & p & 0 \\ 0 & 0 & 0 & s \end{pmatrix} ,~
H2,1=\begin{pmatrix} k^2 & 0 & 0 & 0 \\  0 & kq & 0 & 0 \\ 0 & k^2-pq & kp & 0 \\ 0 & 0 & 0 & k^2 \end{pmatrix} , \nn\\
& & H2,2=\begin{pmatrix} k^2 & 0 & 0 & 0 \\ 0 & kq & 0 & 0 \\ 0 & k^2-pq & kp & 0 \\ 0 & 0 & 0 & -pq \end{pmatrix} ,~
H2,3=\begin{pmatrix} k & p & q & s \\ 0 & k & 0 & q\\ 0 & 0 & k & p \\ 0 & 0 & 0 & k \end{pmatrix}, \nn\\
& & H1,1=\begin{pmatrix} p^2+2pq-q^2 & 0 & 0 & p^2-q^2 \\  0 & p^2+q^2 & p^2 -q^2 & 0 \\ 0 & p^2-q^2 & p^2+q^2 & 0 \\ p^2-q^2 & 0 & 0 & p^2-2pq-q^2 \end{pmatrix}, \nn \\
& & H1,2=\begin{pmatrix} p & 0 & 0 & k \\ 0 & q & 0 & 0 \\ 0 & p-q & p & 0 \\ 0 & 0 & 0 & -q \end{pmatrix}, ~
H1,3 =\begin{pmatrix} k^2 & -kp & kp & pq \\  0 & k^2 & 0 & -kq \\ 0 & 0 & k^2 & kq \\ 0 & 0 & 0 & k^2 \end{pmatrix},~ 
H1,4=\begin{pmatrix} 0 & 0 & 0 & p \\ 0 & 0 & k & 0 \\ 0 & k & 0 & 0 \\ q & 0 & 0 & 0 \end{pmatrix},
\nn \\
& & H0,1=\begin{pmatrix} 1 & 0 & 0 & 1 \\ 0 & -1 & 0 & 0 \\ 0 & 0 & -1 & 0 \\ 0 & 0 & 0 & 1 \end{pmatrix}, ~
H0,2 = \begin{pmatrix} 1 & 0 & 0 & 1 \\  0 & -1 & 1 & 0 \\ 0 & 1 & 1 & 0 \\ -1 & 0 & 0 & 1 \end{pmatrix}. 
\label{eq:Hietarinta4by4}
\end{eqnarray}
These solve the vertex form or the non-braided form of the Yang-Baxter equation. Apart from these solutions we also have the permutation operator $P$ as the eleventh class, which we know can be lifted to higher simplex operators as it satisfies both the required constraints (See Remark \ref{rem:P-YY}).

Both the continuous and discrete symmetries are also satisfied by the two constraint equations as discussed in Sec. \ref{subsec:constraints-d-simplex}. As a consequence of this fact it is sufficient to check if each of the operators in \eqref{eq:Hietarinta4by4} satisfy the YY-constraint. We find that 5 of these classes, represented by $H0,1$, $H0,2$, $H1,4$, $H2,3$ and $H3,1$ satisfy the YY-constraint. 

This can also be seen {\it via} the algebraic representations of these 5 classes as discussed recently in \cite{maity2024algebraic}. From Table 2 of \cite{maity2024algebraic} we know that these 5 classes can be derived from Yang-Baxter solutions obtained using Clifford algebras \cite{padmanabhan2024solving}. More precisely the classes $(0,1)$, and $(1,4)$ can be obtained from 
\begin{eqnarray}
    Y_{jk} = \alpha~A_jA_k + \beta~B_jB_k,~~~\alpha, \beta\in\mathbb{C} \label{eq:YBOanticommuting}
\end{eqnarray}
for different representations of the $A$ and $B$ operators that satisfy
\begin{eqnarray}
    \{A_j, B_j\}=0,~~\left[A_j, B_k\right]=0,~~\textrm{when}~j\neq k.
\end{eqnarray}
The $A$'s and $B$'s need to satisfy further relations realizing two different representations of the above algebra. The details can be found in \cite{maity2024algebraic,padmanabhan2024solving}.
For the $(3,1)$ and $(2,3)$ cases, $A$ and $B$ are orthogonal projectors \cite{maity2024algebraic,padmanabhan2024solving} and hence also commute with each other. The $Y$ operator in this case is given by\begin{eqnarray}
    Y_{jk} & = & \mathbb{1} + \alpha_1~A_j+\alpha_2~A_k + \beta_1~B_j + \beta_2~B_k \nonumber \\
    & + & \gamma_1~A_jA_k + \gamma_2~A_jB_k + \gamma_3~B_jA_k + \gamma_4~B_jB_k, \label{eq:YBOcommuting}
\end{eqnarray}
where the $\alpha$'s, $\beta$'s and $\gamma$'s are complex parameters.
Both of these $Y$ operators satisfy the YY-constraint. While this is readily true for the second $Y$ involving commuting $A$ and $B$, it needs to be verified for the first $Y$ operator in \eqref{eq:YBOanticommuting}. The proof goes as follows:
\begin{eqnarray}
    & & Y_{ij}Y_{kl}Y_{ik}Y_{jl} \nonumber \\
    & = & \alpha~A_jA_l\left(\alpha~A_iA_j-\beta~B_iB_j\right)\left(\alpha~A_kA_l-\beta~B_kB_l\right)\left(\alpha~A_iA_k+\beta~B_iB_k\right) \nonumber \\
    & + & \beta~B_jB_l\left(-\alpha~A_iA_j+\beta~B_iB_j\right)\left(-\alpha~A_kA_l+\beta~B_kB_l\right)\left(\alpha~A_iA_k+\beta~B_iB_k\right) \nonumber \\
    & = & Y_{jl}\left(\alpha~A_iA_j-\beta~B_iB_j\right)\left(\alpha~A_kA_l-\beta~B_kB_l\right)\left(\alpha~A_iA_k+\beta~B_iB_k\right) \nonumber \\
    & = & Y_{jl}\left[\alpha~A_iA_k\left(\alpha~A_iA_j+\beta~B_iB_j\right)\left(\alpha~A_kA_l+\beta~B_kB_l\right) \right.  \nonumber \\ 
    & + & \left. \beta~B_iB_k\left(-\alpha~A_iA_j-\beta~B_iB_j\right)\left(-\alpha~A_kA_l-\beta~B_kB_l\right) \right] \nonumber \\
    & = & Y_{jl}Y_{ik}Y_{kl}Y_{ij}.
\end{eqnarray}
The above proof only requires that $A$ and $B$ anticommute with each other and thus it implies that both the $(0,1)$ and $(1,4)$ class can be lifted to higher simplex operators. 

Finally the $(0,2)$ class is in fact the Majorana braid operator when the Majoranas are realized by the Jordan-Wigner transformation\footnote{This is only true when the nearest-neighbor sites are chosen to label the braid generator. The Majorana representation is non-local and so for indices that are far apart the operator is no longer a 4 by 4 matrix and hence cannot be compared with the Hietarinta solutions. See \cite{maity2024algebraic} for more details.}: 
\begin{equation}
    \gamma_i = X_i \displaystyle \prod_{k=1}^{i-1} Z_k,
\end{equation}
where the $X$ and $Z$ are the first and third Pauli matrices respectively. This satisfies the YY-constraint as shown in Sec. \ref{subsec:majoranafermion}. 

The $M$ matrices corresponding to these 5 Hietarinta classes were studied in detail in \cite{singh2024unitarytetrahedronquantumgates} and so we refer the reader to this reference. Algebraically we note that we can choose the $M$ matrices as either $A$ or $B$ for the Clifford Yang-Baxter solutions. Thus we conclude that the 5 Hietarinta classes $(0,1)$, $(0,2)$, $(1,4)$, $(2,3)$ and $(3,1)$, along with the appropriate $M$ operators \cite{singh2024unitarytetrahedronquantumgates}, can be lifted to higher simplex operators. It is also worth noting that the only solutions among the Hietarinta classes that can be lifted are those that are constructed using Clifford algebra generators \cite{padmanabhan2024solving}.

Two remarks are necessary to complete this analysis.

\begin{remark}
    The remaining 5 classes : $(1,1)$, $(1,2)$, $(1,3)$, $(2,1)$ and $(2,2)$ do not satisfy the YY-constraint and hence cannot be lifted to higher simplex operators\footnote{We can still find compatible $M$ operators, satisfying the YM-constraint and the resulting pair can provide solutions to higher simplex equations when only a single $Y$ operator is involved.}. However in certain regions of their parameter space, describing each of these classes, we can lift them to higher simplex operators. In most of the instances these solutions either become non-invertible or reduce to special cases of the classes that can be lifted. The details are as follows:
    \begin{enumerate}
        \item For the $(1,1)$ class, when either $p=0$ or $q=0$ or $p=q$ we obtain invertible $Y$'s that satisfy the YY-constraint.
        \item For the $(1,2)$ class, when either $p=0$ or $q=0$ we obtain rank 2 matrices that satisfy the YY-constraint. When $k=0$ and $p=q$ this class becomes a special case of the $(1,3)$ class that can be lifted.
        \item For the $(1,3)$ class, the $k=0$ section results in a rank 1 solution that can be lifted. When $q=-p$, this class reduces to a special case of the $(2,3)$ class that can be lifted.
        \item When $k=0$, the $(2,1)$ class becomes a rank 1 matrix that can be lifted. For $p=q=0$ it becomes a rank 3 matrix. When either $k=p=0$ or $k=q=0$ this class reduces to the zero matrix which trivially satisfies the required constraints. Finally when $q=\frac{k^2}{p}$ this class becomes a special case of the $(3,1)$ class that can be lifted.
        \item When $p=0$, the $(2,2)$ class becomes a rank 3 solution that can be lifted. For $k=0$ it becomes a rank 1 operator, and for $q=0$ it becomes a rank 2 solution. When both $k=p=0$ it becomes the zero matrix, a trivial solution. In the last case when $q=\frac{k^2}{p}$ this class becomes a special case of the $(3,1)$ case, which we know can be lifted.
    \end{enumerate}
\end{remark}

\begin{remark}
    It is important to note that the Jones representation \cite{Jones:1985dw}
    \begin{eqnarray}
        Y_{jk} = P_{jk}\left[\alpha + \beta~e_{jk}\right],
    \end{eqnarray}
    does not satisfy the YY-constraint. Here $e$ is the Temperley-Lieb generator parametrized by $\eta$ and $\alpha$, $\beta$ satisfy $\alpha^2+\beta^2+\eta\alpha\beta=0$. This fact can be realized by noticing that 
    $$e_{ij}e_{jk}e_{ik}e_{jl} \neq e_{jl}e_{ik}e_{jk}e_{ij}.$$ Thus such $Y$'s cannot be lifted to higher simplex operators. This is consistent with the fact that the $(2,1)$, $(1,1)$ and $(1,3)$ Hietarinta classes cannot be lifted as these are realized using the Jones representation \cite{maity2024algebraic}. 
\end{remark}

\begin{remark}\label{rem:subclass-1}
    Some of the higher simplex operators obtained by lifting the Clifford Yang-Baxter operators coincide with those found in our previous work \cite{padmanabhan2024solving}. For example, the operator $$Y_{ij}\tilde{Y}_{kl}= \left(\alpha~A_iA_j + \beta~B_iB_j\right)\left(\tilde{\alpha}~A_kA_l + \tilde{\beta}~B_kB_l\right),$$
    results in an operator that is a subclass of the 4-simplex operator constructed out of the $(4,0)$, $(2,2)$ and $(0,4)$ type operators of \cite{padmanabhan2024solving}. Another example involves choosing the Clifford $Y$ and $M_k=A_k$ resulting in a tetrahedron operator that is a subclass of the operators constructed out of types $(3,0)$ and $(1,2)$ \cite{padmanabhan2024solving}.
\end{remark}

\subsection{Summary}
\label{subsec:summary}
The process of lifting Yang-Baxter solutions to higher simplex operators has been successfully carried out for three types of Yang-Baxter operators. The algebras used in these solutions include those of
\begin{enumerate}
    \item Majorana fermions,
    \item Dirac [complex] fermions,
    \item Clifford algebras.
\end{enumerate}
The results are summarized in Table \ref{tab:summary}.

\begin{table}[h!]
   \centering
\begin{tabular}{ |c|c|c|c| }
 \hline
S.No & Method &  $Y$ & $M$  \\
\hline \hline
1. & Majorana fermions & $P_{jk}\left[\mathbb{1} + \gamma_j\gamma_k\right]$  & $\gamma_j$  \\
\hline
2. & Complex fermions - 1 & $\mathbb{1}+\alpha~a_ja_k$ & $\mathbb{1}-2a_j^\dag a_j$  \\
\hline
3. & Complex fermions - 2 & $\mathbb{1}+\alpha~P_{jk}a_ja_k$ & $\mathbb{1}-2a_j^\dag a_j$  \\
\hline
4. & $H0,1$ & $\begin{pmatrix} 1 & 0 & 0 & 1 \\ 0 & -1 & 0 & 0 \\ 0 & 0 & -1 & 0 \\ 0 & 0 & 0 & 1 \end{pmatrix}$ & $Z=\begin{pmatrix}
    1 & 0 \\ 0 & -1
\end{pmatrix}$  \\
\hline
5. & $H0,2$ & $\begin{pmatrix} 1 & 0 & 0 & 1 \\  0 & -1 & 1 & 0 \\ 0 & 1 & 1 & 0 \\ -1 & 0 & 0 & 1 \end{pmatrix}$ & $Z=\begin{pmatrix}
    1 & 0 \\ 0 & -1
\end{pmatrix}$ \\
\hline
6. & $H1,4$ & $\begin{pmatrix} 0 & 0 & 0 & p \\ 0 & 0 & k & 0 \\ 0 & k & 0 & 0 \\ q & 0 & 0 & 0 \end{pmatrix}$ & $\left\{\begin{pmatrix} 1 & 0  \\ 0 & -1  \end{pmatrix}, \begin{pmatrix} 0& 1 \\ \pm \frac{\sqrt{q}}{\sqrt{p}} & 0  \end{pmatrix}\right\}$ \\
\hline
7. & $H2,3$ & $\begin{pmatrix} k & p & q & s \\ 0 & k & 0 & q\\ 0 & 0 & k & p \\ 0 & 0 & 0 & k \end{pmatrix}$ & $\begin{pmatrix} m_1 & m_2  \\ 0 & m_4  \end{pmatrix}~m_1, m_2, m_4\in\mathbb{C}$ \\
\hline
8. & $H3,1$ & $\begin{pmatrix} k & 0 & 0 & 0 \\ 0 & q & 0 & 0 \\ 0 & 0 & p & 0 \\ 0 & 0 & 0 & s \end{pmatrix}$ & $\begin{pmatrix} m_1 & 0  \\ 0 & m_4  \end{pmatrix}~m_1,m_4\in\mathbb{C}$\\
\hline
\end{tabular}
\caption{The different types of Yang-Baxter operators that can be lifted to higher simplex operators. The $\gamma$'s and $a$'s refer to Majorana and complex fermions respectively. While the first three entries are algebraic solutions, the remaining are the two dimensional solutions of Hietarinta. Their algebraic versions can be found in \eqref{eq:YBOanticommuting} and \eqref{eq:YBOcommuting}. The algebraic versions of the corresponding $M$ operators are given by either $A_j$ or $B_j$ for the Yang-Baxter operator in \eqref{eq:YBOanticommuting} and $\left(\mathbb{1}+k_1A_j+k_2B_j+k_3A_jB_j+k_4B_jA_j\right)$ for the Yang-Baxter operator in \eqref{eq:YBOcommuting}. When the two dimensional representation is chosen, the Majorana solution is equivalent to $H0,2$ and the two fermion solutions are a special case of the $(2,3)$ class.}
\label{tab:summary}
\end{table}

\section{Positivity of Boltzmann weights}
\label{sec:positivity}
Before we extend the formalism of lifting to other lower simplex operators, we will analyze the different applications of the solutions obtained so far. The Yang-Baxter solutions constructed using Majorana fermions are most relevant for studying the braiding properties of Majorana fermions as seen in \cite{Ivanov_2001}. Subsequently they also made an appearance in fermionic quantum computation \cite{Bravyi_2002} and much later they found applications in topological quantum computation \cite{kauffman2013knotlogictopologicalquantum,Kauffman_majorana, Kauffman_2016} as quantum gates. We expect similar applications for the Majorana fermion higher simplex operators obtained by lifting this Majorana Yang-Baxter solution. These require a more detailed and careful study which we postpone to a future work. Nevertheless, we take a brief look at some of the braiding properties of the Majorana tetrahedron solutions in Sec. \ref{sec:conclusion}.

The other application of the solutions seen up to this point, is in the construction of integrable quantum systems in different spacetime dimensions. This is well understood in the case of $1+1$ dimensions, where the Yang-Baxter equation leads to a tower of conserved quantities, that also includes, in many cases, a local Hamiltonian. The partition function of such Hamiltonians can be found using a state-sum construction, where the local constituents are interpreted as local Boltzmann weights which are in turn functions of the matrix elements of the Yang-Baxter solutions. A similar construction is desirable for the tetrahedron and other higher simplex solutions to help build integrable quantum models in higher dimensions. While a tower of commuting operators and a Hamiltonian in two and higher dimensions using these solutions would be quite remarkable, we do not attempt that in this work. We can still study the partition functions of such higher dimensional integrable systems by adapting the state-sum construction used in the Yang-Baxter ($1+1$) case. In this setting the local Boltzmann weights are functions of the matrix elements of the tetrahedron operators in the three dimensional case, and other higher simplex operators in the associated dimension. To obtain physically relevant examples we require that these Boltzmann weights are positive. A sufficient condition to fulfill this is when the matrix elements of the tetrahedron operators are themselves positive. This has often been a challenge in the past. We will now check which of the solutions obtained so far have this property. For this purpose the Clifford solutions shown in Table \ref{tab:summary} are the most appropriate ones.

Before we study the $2+1$ dimensional case we briefly review the state-sum construction of the partition function of integrable systems in $1+1$ dimensions and see how this generalizes to the higher dimensional case. We start by embedding a fixed number of oriented straight lines in Euclidean $\mathbb{R}^2$. Physically these are the world lines of particles moving on a straight line. There are many such embeddings but we consider particular configurations which satisfy the conditions:
\begin{enumerate}
    \item No three lines intersect at a point.
    \item No two lines are parallel.
\end{enumerate}
These configurations are such that each line is divided into several segments due to the intersection with the other lines. All except two of those segments remain bounded by the points of intersection. The {\it state} $\sigma$ of a configuration is when each segment is attached a label from a set $\{1,2,\cdots, n\}$, called a {\it color}. The integer $n$ is typically the dimension of the local Hilbert space on which the Yang-Baxter operator $Y$ acts, $$ Y : \mathbb{C}^n\otimes\mathbb{C}^n\rightarrow \mathbb{C}^n\otimes\mathbb{C}^n.$$ We can then define the {\it Boltzmann weight} $E(\sigma)$ as 
\begin{eqnarray}
    E(\sigma) = \prod\limits_v~E_v(\sigma) = \prod\limits_v~Y_{ij}^{kl},
\end{eqnarray} where the product runs over all the vertices of the configuration. The indices $\{i,j ; k, l\}$ are the incoming and outgoing colors on the segments surrounding the vertex $v$ as shown in the figure. 
\begin{center}
  \begin{tikzpicture}
    \draw[->, thick] (-1.2,0) -- (1.2,0) node[right] {\( k \)};
    \draw[->, thick] (0,-1.2) -- (0,1.2) node[above] {\( l \)};
    \fill (0,0) circle (2pt);
    \node[below left] at (0,0) {\( v \)};
    \node[left] at (-1.2,0) {\( i \)};
    \node[below] at (0,-1.2) {\( j \)};
\end{tikzpicture}  
\end{center}
The Boltzmann weight is the exponential of the energy function and so we expect it to be positive. From the above expression it is clear that a sufficient condition for this is when the matrix entries $Y_{ij}^{kl}$ are positive. Note that this is not a necessary condition. Using these Boltzmann weights the partition function of the system is given by 
\begin{eqnarray}
    Z = \sum\limits_{\sigma}~E(\sigma).
\end{eqnarray} 
It is important to note that the above discussion was for constant Yang-Baxter operators without spectral parameters. This construction can be easily repeated for those with spectral parameters as well. In that case the colors on the segments are also functions of the complex spectral parameters. The functions $E(\sigma)_v$ depend on these spectral parameters and they are interpreted as slopes of the straight lines intersecting at the vertex $v$. 

This construction can be repeated in the $2+1$ case with the straight lines in $\mathbb{R}^2$ replaced by ordered and oriented planes embedded in $\mathbb{R}^3$. Among the different possible configurations we only consider those where the intersection of $i$ planes has codimension exactly $i$, with $i=2,3,4$. The resulting configurations, thus have vertices, segments and faces, obtained at the intersection of three planes, two planes and one plane respectively. These configurations split $\mathbb{R}^3$ into {\it regions}. We can now define different statistical mechanical models by coloring either the regions, faces or segments. As a consequence the tetrahedron equation is also not unique and takes different forms. We will not go into the details here, but will use the segment coloring to arrive at the vertex form of the tetrahedron equation. In this case the Boltzmann weights are made out of 
\begin{eqnarray}
    E_v(\sigma) = T_{i_{ab}, i_{ac}, i_{bc} }^{j_{ab}, j_{ac}, j_{bc} },
\end{eqnarray} where $a$, $b$ and $c$ are the labels of the planes intersecting at the point $v$ and $T$ is the tetrahedron solution. The indices $\{ab, ac, bc\}$ label the segments formed at the intersection of the corresponding planes. The colors are once again attached to the segments coming out of the vertex $v$. As before we expect physically meaningful expressions when the matrix elements of the tetrahedron operators are positive leading to positive Boltzmann weights.

All of the solutions displayed in Table \ref{tab:summary} solve the constant versions of the tetrahedron and other higher simplex equations. A sufficient condition for positive Boltzmann weights is by ensuring that the entries of the $Y$ and $M$ operators are themselves positive. We see that to be the case in three of the $Y$, $M$ pairs from Table \ref{tab:summary}, namely the $H1,4$, $H2,3$ and the $H3,1$ classes. More precisely:
\begin{enumerate}
    \item For the $H1,4$ class we choose the parameters $p$, $k$ and $q$ in the $Y$-matrix to be positive and for the $M$ matrix we choose $\begin{pmatrix}
    0 & 1 \\ \frac{\sqrt{q}}{\sqrt{p}} & 0
\end{pmatrix}$.
\item Choose positive parameters for all parameters in $Y$ and $M$ for the $H2,3$ class.
\item Choose positive parameters for all parameters in $Y$ and $M$ for the $H3,1$ class.
\end{enumerate}
This completes the analysis for the constant solutions. The analysis can be extended to the spectral parameter dependent case as well.
The Clifford solutions in \eqref{eq:YBOanticommuting} can easily be Baxterized to a non-regular solution of the Yang-Baxter equation by converting the constants $\alpha$ and $\beta$ to site-dependent parameters, $\alpha_{jk}$ and $\beta_{jk}$ respectively,
\begin{eqnarray}
 Y_{jk}(\alpha_{jk},\beta_{jk}) = \alpha_{jk}~A_jA_k + \beta_{jk}~B_jB_k~;~\alpha_{jk}, \beta_{jk}\in\mathbb{C}.   \end{eqnarray}
They satisfy the spectral parameter dependent Yang-Baxter equation 
\begin{eqnarray}
    Y_{ij}(\alpha_{ij},\beta_{ij})Y_{ik}(\alpha_{ik},\beta_{ik})Y_{jk}(\alpha_{jk},\beta_{jk}) = Y_{jk}(\alpha_{jk},\beta_{jk})Y_{ik}(\alpha_{ik},\beta_{ik})Y_{ij}(\alpha_{ij},\beta_{ij}),
\end{eqnarray}
with the spectral parameters appearing in non-additive form. Such non-regular solutions were also written in a recent work on the classification of non-regular Yang-Baxter operators \cite{deleeuw20254x4solutions} (See Eqs. 22, 23, and 24 there). It is easy to check that these operators also satisfy the $YY$ and the $YM$ constraints with constant $M$ operators and can thus be lifted to tetrahedron and other higher simplex solutions. By appropriate choices of the spectral parameters we end up with positive Boltzmann weights for these cases as well.

\section{Lifting the tetrahedron operator}
\label{sec:tetrahedron}
So far we have used the 1-simplex and 2-simplex operators $M$ and $Y$, subject to the YM- and YY-constraints \eqref{eq:YMMconstraint-1} and \eqref{eq:YYconstraint}, to construct the higher $d$-simplex solutions. This amounts to the different ways of partitioning the integer $d$ into 1's and 2's. We have also noted that for the YM- and YY-constraints to hold, we further require that the two indices on the $Y$ operators have to be nearest neighbors .  

Now we will discuss the constraints that arise when we also include a 3-simplex (tetrahedron) operator $T$ into this mix. Besides the two constraints from the $Y$ and $M$ operators, we expect three more constraints in this situation. The first one involving $T$ and $M$, the second one with $T$ and $Y$ and a third one just among the $T$'s. The latter is analogous to the $YY$-constraint. 

There is a systematic way to obtain these additional constraints as we shall now see. To obtain the constraint between the $T$ and $M$ operators we substitute the ansatz 
\begin{eqnarray}\label{31to4simplex}
    R_{ijkl} = T_{ijk}M_l,
\end{eqnarray}
into the 4-simplex equation \eqref{eq:4-simplex}. Then the ansatz solves the latter when 
\begin{eqnarray}\label{eq:TMMMconstraint-1}
    M_iM_jM_kT_{ijk} = T_{ijk}M_kM_jM_i,
\end{eqnarray}
which we denote as the {\it TM-constraint}. When the $M$'s on different sites commute with each other this constraint can be written as 
\begin{eqnarray}\label{eq:TMMMconstraint-2}
   \left[T, M\otimes M\otimes M\right] = 0. 
\end{eqnarray}
The constraint between the $T$ and $Y$ operators is obtained by requiring that the operator 
\begin{eqnarray}\label{eq:32to5simplex}
    R_{ijklm} = T_{ijk}Y_{lm},
\end{eqnarray}
satisfies the 5-simplex equation
\begin{eqnarray}
    R_{ijklm}R_{inpqr}R_{jnstu}R_{kpsvx}R_{lqtvy}R_{mruxy} = R_{myuxy}R_{lqtvy}R_{kpsvx}R_{jnstu}R_{inpqr}R_{ijklm}. \nonumber \\
    \label{eq:5-simplex}
\end{eqnarray}
The resulting constraint, denoted the {\it TY-constraint}, is given by 
\begin{eqnarray}\label{eq:TYconstraint}
    Y_{ij}Y_{kl}Y_{mn}T_{ikm}T_{jln} = T_{jln}T_{ikm}Y_{mn}Y_{kl}Y_{ij}.
\end{eqnarray}
Lastly we obtain the constraint among the $T$ operators by plugging in the ansatz
\begin{eqnarray}
    R_{ijkuvw}= T_{ijk}T_{uvw}
\end{eqnarray}
into the 6-simplex equation 
\begin{eqnarray}
    & & R_{ijkuvw}R_{ilmnxy}R_{jlabcd}R_{kmaefg}R_{unbehp}R_{vxcfhq}R_{wydgpq} \nonumber \\
    & = & R_{wydgpq}R_{vxcfhq}R_{unbehp}R_{kmaefg}R_{jlabcd}R_{ilmnxy}R_{ijkuvw}.
\end{eqnarray}
The {\it TT-constraint} becomes 
\begin{eqnarray}\label{eq:TTconstraint}
    T_{ijk}T_{uvw}T_{xyz}T_{iux}T_{jvy}T_{kwz} = T_{kwz}T_{jvy}T_{iux}T_{xyz}T_{uvw}T_{ijk}.
\end{eqnarray}
Besides these three constraints, we require that the indices on the $T$ and $Y$ operators are `nearest neighbors' (See Sec. \ref{subsec:constraints-d-simplex} and Remark \ref{rem:forbiddenYM}). As an example, for a 5-simplex operator $R_{ijklm}$ splitting into a $T$ and $Y$, a valid choice is $Y_{ij}T_{klm}$ and a forbidden choice is $T_{ilm}Y_{jk}$.

Thus to lift the tetrahedron (3-simplex) operators along with the $Y$ (2-simplex) and $M$ (1-simplex) operators we require that they satisfy 5 constraints, the YM-, YY-, TM, TY- and TT-constraints. On top of this the indices on the $Y$ and $T$ operators have to be nearest neighbors. 

Note that we do not obtain a constraint involving all three, $T$, $Y$ and $M$ operators. This can be understood from the fact that the every index appears exactly twice in the vertex form of the tetrahedron equation \eqref{eq:3-simplex}. We will use this fact to generalize the constraint analysis while lifting arbitrary $d$-simplex operators. 

\begin{remark}
    Just as the Yang-Baxter equation, the tetrahedron equation has both continuous and discrete symmetries. These symmetries carry over to the TM-, TY- and TT-constraints as well. Note that all three operators $M$, $Y$ and $T$ have to be conjugated by the same $Q$ operator for the continuous symmetries to hold. 
\end{remark}

We will now construct the solutions satisfying these constraints. As in the lifting of the Yang-Baxter operators we use the generators of three algebras : Majorana and Dirac (complex) fermions, and a pair of Clifford algebra generators. We find that only the latter two provide tetrahedron operators that can be lifted consistently while the former provides solutions that cannot be lifted using the procedures outlined in this work.

\paragraph{Dirac (complex) fermions : } The operator set 
\begin{eqnarray}\label{eq:TYMfermions}
    M_j = \mathbb{1} -2a_j^\dag a_j~;~Y_{jk} = \mathbb{1} + \mu~a_ja_k~;~T_{ijk} = \mathbb{1} + \alpha~a_ia_j + \beta~a_ja_k +\delta~a_ka_i.
\end{eqnarray}
Here $\mu$, $\alpha$, $\beta$ and $\delta$ are complex parameters. The lifting of the $Y$ and $M$ operators are studied in Sec. \ref{subsec:complexfermion}. With a few lines of computation it can be verified that they satisfy the TT-, TM- and the TY-constraints. In particular the constraints immediately follow from the identities 
$$ T_{ijk}\tilde{T}_{iux} = \tilde{T}_{iux}T_{ijk}~;~Y_{ij}T_{ikm} = T_{ikm}Y_{ij},$$
with $\tilde{T}_{iux} = \mathbb{1} + \tilde{\alpha}~a_ia_j + \tilde{\beta}~a_ja_k +\tilde{\delta}~a_ka_i $.

The above solution set is invertible. A non-invertible set is obtained by replacing the above $M_j$ operator with $a_j$. Apart from these solutions it is possible to obtain $T$ operators that cannot be lifted using the methods discussed here. Such an example is given by 
\begin{eqnarray}
    T_{ijk} = \mathbb{1} + \alpha~a_ia_kja_k~;~\alpha\in\mathbb{C}.
\end{eqnarray}

\begin{remark}
    It should be noted that the tetrahedron operator in \eqref{eq:TYMfermions} can be obtained from $Y$ operators as follows
    \begin{eqnarray}
        T_{ijk} = Y^\alpha_{ij}Y^\beta_{jk}Y^\delta_{ki}~;~Y_{ij}^{\mu} = \mathbb{1} + \mu~a_ia_j.
    \end{eqnarray}
    However these satisfy a different set of YY-constraints from \eqref{eq:YYconstraint} to arrive at the tetrahedron solution. See also \cite{CarterSaito} for a method to construct tetrahedron operators purely out of Yang-Baxter operators. However in these methods the Hilbert spaces get doubled.
\end{remark}

\paragraph{Clifford algebras :} We use a pair of anticommuting operators $A$ and $B$, that can be realized using Clifford algebras \cite{padmanabhan2024solving}, to construct this set. Consider the operators
\begin{eqnarray}\label{eq:TYMclifford}
    M_j & = & A_j, \nonumber \\
    Y_{jk} & = & \mu~A_jA_k + \nu~B_jB_k,~~\mu,\nu\in\mathbb{C}, \nonumber \\
    T_{ijk} & = & \alpha~A_iA_jA_k + \beta_1~B_iB_jA_k + \beta_2~B_iA_jB_k + \beta_3~A_iB_jB_k,~~\alpha, \beta_j\in\mathbb{C}.
\end{eqnarray}
In the notation of \cite{padmanabhan2024solving}, the $Y$ operator is made of the $(2,0)$ and $(0,2)$ operator types, whereas the $T$ operator is a linear combination of the $(3,0)$, $(1,2)$ operator types. By interchanging $A$ and $B$ in the above expressions we obtain another set of solutions which are not equivalent to the above set of operators. 

We have seen in Sec. \ref{subsec:hietarinta} that the $Y$ and $M$ operators satisfy the YM- and the YY-constraints. We are left with the verification of the TM-, TY-, and TT-constraints. While the first one is relatively simpler the latter two require some work which we briefly outline below. For the TY-constraint we have 
\begin{eqnarray}
    && Y_{ij}Y_{kl}Y_{mn}T_{ikm}T_{jln} \nonumber \\
    & = & T_{ikm}\left(\mu A_iA_j-\nu B_iB_j \right)\left(\mu A_kA_l-\nu B_kB_l \right)\left(\mu A_mA_n-\nu B_mB_n \right)T_{jln} \nonumber \\
    & = & T_{ikm}T_{jln}Y_{ij}Y_{kl}Y_{mn}\nonumber \\
    & = & T_{jln}T_{ikm}Y_{mn}Y_{kl}Y_{ij}.
\end{eqnarray}
A similar proof goes through for the TT-constraint as 
\begin{eqnarray}
    && T_{ijk}T_{uvw}T_{xyz}T_{iux}T_{jvy}T_{kwz} \nonumber \\
    & = & T_{iux}\left(+,-,-,+\right)_{ijk}\left(+,-,-,+\right)_{uvw}\left(+,-,-,+\right)_{xyz}T_{jvy}T_{kwz} \nonumber \\
    & = & T_{iux}T_{jvy}\left(+,+,-,-\right)_{ijk}\left(+,+,-,-\right)_{uvw}\left(+,+,-,-\right)_{xyz}T_{kwz} \nonumber \\
    & = & T_{ikx}T_{jvy}T_{kwz}T_{ijk}T_{uvw}T_{xyz} \nonumber \\
    & = & T_{kwz}T_{jvy}T_{ikx}T_{xyz}T_{uvw}T_{ijk}.
\end{eqnarray}
To keep things succinct we have used the following notation \cite{padmanabhan2024solving}
$$ \left(s_1,s_2,s_3,s_4\right)_{ijk} = s_1~\alpha A_iA_jA_k + s_2~\beta_1B_iB_jA_k + s_3~\beta_2B_iA_jB_k + s_4~\beta_3A_iB_jB_k, $$
with $s_j\in\{+,-\}$, for the tetrahedron operators.  

\begin{remark}\label{rem:subclass-2}
    For the 4-simplex case we observed that some of the $YY$ solutions result in 4-simplex operators that are subclasses of the 4-simplex operators constructed in \cite{padmanabhan2024solving}. Similar statements hold for the $TY$ and $TT$ solutions here. They result in operators that can be seen as subclasses of the 5- and 6-simplex operators constructed in \cite{padmanabhan2024solving}.
\end{remark}

\paragraph{Majorana fermions : } Unlike the previous two cases there seems to be no tetrahedron solution constructed using Majorana fermions that can be lifted to higher simplex solutions. Nevertheless we do find interesting Majorana solutions for the tetrahedron operator 
\begin{eqnarray}\label{eq:Tmajorana}
    T_{ijk} = \mathbb{1} + \alpha~\gamma_i\gamma_j\gamma_k~;~\alpha\in\mathbb{C}.
\end{eqnarray}
It is easy to see that this operator easily solves the tetrahedron equation as any two terms of this equation share a single index and hence the above operators commute with each other, thus satisfying the tetrahedron equation. This logic holds for all $2d+1$-simplex equations, implying that the operator
\begin{eqnarray}
    R_{i_1\cdots i_{2d+1}} = \mathbb{1} + \alpha~\gamma_{i_1}\cdots\gamma_{i_{2d+1}},
\end{eqnarray}
is a $2d+1$-simplex operator. These operators are invertible with the inverse given by 
\begin{eqnarray}
  R_{i_1\cdots i_{2d+1}}^{-1} = \frac{1}{1 + (-1)^d\alpha^2} ~\left[\mathbb{1} - \alpha~\gamma_{i_1}\cdots\gamma_{i_{2d+1}}\right]. 
\end{eqnarray}

\section{Lifting arbitrary $d$-simplex operators}
\label{sec:generalLifts}
The analysis of the lifting of 2- and 3-simplex operators helps us write down the procedure for the lifting of an arbitrary $d$-simplex operator. We begin with a set of $j$-simplex operators, for $j\in\{1,\cdots, d\}$. Then a $h$-simplex operator, for $h>d$, can be constructed using these in all the ways the integer $h$ can be partitioned into the $j$-simplex operators. More precisely we have
\begin{eqnarray}
    h = \sum\limits_{j=1}^{d}~n_jj,
\end{eqnarray}
where $n_j$ is the number of times the $j$-simplex operator appears in the partition. Substituting this operator into the vertex form of the $h$-simplex equation reveals a number of constraints on the $j$-simplex operators that need to be satisfied for the lifting to succeed. As every index appears precisely twice in the vertex form of the $h$-simplex equation we expect constraints only among any two types of lower simplex operators. This statement requires a proof which we do not provide here. However this has been verified in a number of cases (the 2- and 3-simplex cases are shown here). We find that between two different types of $j$-simplex operators there are $\frac{d(d-1)}{2}$ constraints and the number of constraints satisfied by the $j$-simplex operator with itself, sum to $d-1$. Thus the total number of constraints that need to be satisfied for successfully lifting a $d$-simplex operator is given by 
\begin{eqnarray}
    \frac{d(d-1)}{2} + d-1 = \frac{(d-1)(d+2)}{2}.
\end{eqnarray}
This is verified for the cases studied in this work, $d=2$ and $d=3$, we obtain 2 and 5 constraints respectively. 

The form of these constraints is also easily written down by observing the pattern emerging from the cases studied earlier. To understand this let us look at the constraint of a given $j$-simplex operator with itself. The underlying pattern can be seen in the TT-constraint \eqref{eq:TTconstraint} or YY-constraint \eqref{eq:YYconstraint}. The number of operators appearing on each side of this constraint is just $2j$. Let the $j$-simplex operator be denoted $R^{(j)}$, then the $R^{(j)}R^{(j)}$-constraint becomes 
\begin{eqnarray}\label{eq:jjconstraint}
\left[\prod\limits_{k=1}^{j}~R^{(j)}_{\alpha^{(k)}_1\cdots\alpha^{(k)}_j}\right]\left[\prod\limits_{k=1}^{j}~R^{(j)}_{\alpha^{(1)}_k\cdots\alpha^{(j)}_k} \right] = \left[\prod\limits_{k=1}^{j}~R^{(j)}_{\alpha^{(1)}_k\cdots\alpha^{(j)}_k} \right] \left[\prod\limits_{k=1}^{j}~R^{(j)}_{\alpha^{(k)}_1\cdots\alpha^{(k)}_j}\right].
\end{eqnarray}
Barring the 1-simplex operator, there are $(d-1)$ such constraints. Next we write down the constraint between two different types of lower simplex operators, say $j$- and a $l$-simplex operator, denoted by $R^{(j)}$ and $R^{(l)}$ respectively. We expect $l$ $j$-simplex operators and $j$ $l$-simplex operators on each side of this constraint. The index structure is easily gleaned of the TY-constraint \eqref{eq:TYconstraint}. We have 
\begin{eqnarray}\label{eq:ljconstraint}
\left[\prod\limits_{k=1}^{l}~R^{(j)}_{\alpha^{(k)}_1\cdots\alpha^{(k)}_j}\right]\left[\prod\limits_{k=1}^{j}~R^{(l)}_{\alpha^{(1)}_k\cdots\alpha^{(l)}_k} \right] = \left[\prod\limits_{k=1}^{j}~R^{(l)}_{\alpha^{(1)}_k\cdots\alpha^{(l)}_k} \right]\left[\prod\limits_{k=1}^{l}~R^{(j)}_{\alpha^{(k)}_1\cdots\alpha^{(k)}_j}\right].
\end{eqnarray}
For $j\neq l$ we find $\frac{d(d-1)}{2}$ such constraints. Note that when $j=l$, the $jl$-constraint \eqref{eq:ljconstraint} reduces to the $jj$-constraint \eqref{eq:jjconstraint}. 

As for the operators satisfying these constraints, we expect the Clifford solutions of \cite{padmanabhan2024solving} to obey them. Thus all such solutions can be lifted. This is also evident from the fact that many of the solutions constructed here form subclasses of the solutions obtained in \cite{padmanabhan2024solving} (See Remarks \ref{rem:subclass-1} and \ref{rem:subclass-2}). On the other hand the solutions constructed using Majorana and Dirac fermions require separate analysis which we reserve for the future.

\section{Lifting anti-Yang-Baxter solutions}
\label{sec:lifting-antiYBO}
Besides the above solutions we can obtain more higher simplex solutions by lifting anti-lower simplex solutions. The latter solve the anti-lower simplex equations. Their study warrants a separate work, nevertheless we briefly demonstrate these ideas by lifting the anti-Yang-Baxter solutions.

The anti-Yang-Baxter equation was introduced in \cite{padmanabhan2024solving}. Denote their solutions as $Y_{ij}$, to be consistent with the notation of the current paper. This operator satisfies
\begin{eqnarray}\label{eq:anti-YBE-constant}
    Y_{ij}Y_{ik}Y_{jk} = - Y_{jk}Y_{ik}Y_{ij}.
\end{eqnarray}
This is the constant version of the more general 
\begin{eqnarray}\label{eq:anti-YBE-spec}
    Y_{ij}(u_i, u_j)Y_{ik}(u_i, u_k)Y_{jk}(u_j, u_k) = - Y_{jk}(u_j, u_k)Y_{ik}(u_i, u_k)Y_{ij}(u_i, u_j).
\end{eqnarray}
Here the $u$'s are complex spectral parameters. For monodromy matrices constructed out of such $Y$'s conserved quantities appear on lattices with an even number of sites. This result can be seen as a generalization of Theorem 1, Chapter 6 of \cite{Korepin1993QuantumIS}. Thus anti-Yang-Baxter equations and their solutions can also be used to construct integrable models in two dimensions.

The anti-Yang-Baxter operators can also be lifted provided they satisfy a new set of constraints that are the `anti-' versions of the $YM$ and $YY$ (\eqref{eq:YY-YM-constraints}). We will see how the constraints are modified through the examples. As in the rest of this paper, the solutions are obtained using Majorana and Dirac fermions and Clifford algebras. 
\paragraph{From Majorana fermions : } Consider 
\begin{eqnarray}\label{eq:anti-Majorana-YM}
    Y_{ij} = P_{ij}\left[\gamma_i + \gamma_j\right]~;~M=\gamma.
\end{eqnarray} This $Y$ satisfies the anti-Yang-Baxter equation in \eqref{eq:anti-YBE-constant}. Additionally, together with $M$, the operators satisfy the constraints
\begin{eqnarray}
    & M_iM_jY_{ij} = Y_{ij}M_jM_i & \nonumber \\
    & Y_{ij}Y_{kl}Y_{ik}Y_{jl} = - Y_{jl}Y_{ik}Y_{kl}Y_{ij}.    &
\end{eqnarray}
The inverse of these operators are given by
\begin{eqnarray}
    Y_{ij}^{-1} = \frac{1}{2}\left[\gamma_i + \gamma_j\right]P_{ij}~;~M^{-1}=\gamma.
\end{eqnarray}
Using these we can construct the tetrahedron operator
\begin{eqnarray}\label{eq:anti-Majorana}
    T_{ijk} = P_{ij}\left[\gamma_i + \gamma_j\right]\gamma_k
\end{eqnarray}
and the 5-simplex operator
\begin{eqnarray}\label{eq:anti-Majorana-5-simplex}
    R_{ijklm} = P_{ij}\left[\gamma_i + \gamma_j\right]P_{kl}\left[\gamma_k + \gamma_l\right]\gamma_m.
\end{eqnarray}
The 4-indexed operator $R_{ijkl}=P_{ij}\left[\gamma_i + \gamma_j\right]P_{kl}\left[\gamma_k + \gamma_l\right]$ satisfies the anti-4-simplex equation.

The above operators solve the constant higher simplex equations. We do not have expressions for the spectral parameter dependent solutions at this point. Note also that the Majorana solutions find natural applications as braid operators and in topological quantum computation. In light of this, the braiding features of the higher simplex operators constructed out of Majorana fermions becomes important. This will be mentioned in Sec. \ref{sec:conclusion}.
\paragraph{From Dirac(complex) fermions : } The operator\footnote{Interestingly this operator also satisfies the braided form of the Yang-Baxter equation $$ Y_{ij}Y_{jk}Y_{ij} = Y_{jk}Y_{ij}Y_{jk}, $$ in a curious manner as it reduces both sides of the equation to 0. Due to this feature any higher simplex operator constructed out of this $Y$ will solve the corresponding constant higher simplex equation regardless of the choice of $M$. In other words the constraints do not matter while lifting this $Y$. However we should note that the resulting constant higher simplex solutions are all non-invertible.}
\begin{eqnarray}\label{eq:anti-DiracY}
    Y_{ij} = a_i + \alpha~a_j~;~\alpha\in\mathbb{C},
\end{eqnarray} satisfies the constant anti-Yang-Baxter equation in \eqref{eq:anti-YBE-constant}.
The $YY$-constraint \eqref{eq:YYconstraint} also checks out with this operator. We find that for the $Y$ in \eqref{eq:anti-DiracY}, the $YY$-constraint \eqref{eq:YYconstraint} becomes,
\begin{eqnarray}
    Y_{ij}Y_{kl}Y_{ik}Y_{jl} = Y_{jl}Y_{ik}Y_{kl}Y_{ij}=0.
\end{eqnarray} 
Due to this identity, all higher simplex operators constructed with at least two such $Y$'s will solve the corresponding higher simplex equation regardless of the choice of the $M$ operator. In such cases, the $Y$ and $M$ need not satisfy any constraint.

Furthermore, the $M$ operator given by 
\begin{eqnarray}
    M = a,
\end{eqnarray}
satisfies the $YM$-constraint \eqref{eq:YMMconstraint-1} as follows
\begin{eqnarray}
    M_iM_jY_{ij} = Y_{ij}M_jM_i =0.
\end{eqnarray} As a result of this identity, any higher simplex operator containing at least one of the $M$ in its factors, satisfies the corresponding higher simplex equation.

In both the above cases the higher simplex solutions are all non-invertible as the $Y$ in \eqref{eq:anti-DiracY} is non-invertible. Nevertheless we can construct invertible solutions by Baxterizing the $Y$ in \eqref{eq:anti-DiracY}. The operator
\begin{eqnarray}\label{eq:anti-specY-Dirac}
    Y_{ij}(u) = \mathbb{1} +cu~(a_i+\alpha~a_j)~;~c,u\in\mathbb{C},
\end{eqnarray} 
satisfies the additive form of the spectral-parameter dependent Yang-Baxter equation in the braided form,
\begin{eqnarray}
    Y_{ij}(u)Y_{jk}(u+v)Y_{ij}(v) = Y_{jk}(v)Y_{ij}(u+v)Y_{jk}(u).
\end{eqnarray}
This spectral parameter dependent operator needs to be lifted into spectral parameter dependent higher simplex solutions. However, these solutions do not satisfy a spectral parameter dependent version of the $YY$-constraint and thus cannot be lifted to higher simplex solutions containing more than one $Y$ operator. Moreover, the only $M$ operator that this solution accommodates is the trivial identity operator, that helps satisfy a spectral parameter dependent version of the $YM$-constraint. Thus we do not obtain non-trivial higher simplex solutions that consist of both the $Y$ and $M$ operators as well. These considerations do not rule out the possibility of obtaining invertible higher simplex solutions obtained from other Baxterized forms of the $Y$ operator, than the one shown in \eqref{eq:anti-specY-Dirac}. We leave this analysis for a future work. 
\paragraph{From Clifford algebras : }Next we list some Clifford solutions constructed out of two anticommuting operators $A$ and $B$. The operator
\begin{eqnarray}\label{eq:anti-CliffordY}
    Y_{ij} = \alpha~A_iB_j + \beta~B_iA_j~;~\alpha, \beta\in\mathbb{C},
\end{eqnarray}
satisfies the anti-Yang-Baxter equation as shown in \cite{padmanabhan2024solving}. Together with $M=A$ or $M=B$, they satisfy the constraints
\begin{eqnarray}
    & M_iM_jY_{ij} = -Y_{ij}M_jM_i & \nonumber \\
    & Y_{ij}Y_{kl}Y_{ik}Y_{jl} =  Y_{jl}Y_{ik}Y_{kl}Y_{ij}.&
\end{eqnarray}
The higher simplex operators obtained from tensoring these $Y$'s and $M$'s satisfy either the corresponding higher simplex equation or the anti version of the same. This depends on the number of the times, the two constraints and the anti-Yang-Baxter equation appear in the higher simplex equation. For example in the tetrahedron case, the operator $T_{ijk}=Y_{ij}M_k$ results in a single occurrence of both the anti-Yang-Baxter equation and the $YM$-constraint. Thus this operator satisfies the {\it regular} tetrahedron equation, that is without the overall negative sign on the right hand side of the equation. In a similar manner the 4-simplex operator $Y_{ij}Y_{kl}$ and the 5-simplex operator $Y_{ij}Y_{kl}M_n$, satisfy the respective higher simplex equations.  
Moreover the higher simplex operators constructed out of these solutions are seen to be subclasses of the solutions obtained in \cite{padmanabhan2024solving}.

The above solutions for $Y$ and $M$ provide constant higher simplex operators (without spectral parameters) upon lifting. This is easily generalized to the spectral parameter dependent case with the observation that the operator
\begin{equation}
    Y_{ij}(\alpha_{ij}, \beta_{ij})=\alpha_{ij}~A_iB_j + \beta_{ij}~B_iA_j~;~\alpha_{ij}, \beta_{ij}\in\mathbb{C},
\end{equation}
satisfies the non-additive form of the spectral parameter dependent anti-Yang-Baxter equation
\begin{eqnarray}
    Y_{ij}(\alpha_{ij}, \beta_{ij})Y_{ik}(\alpha_{ik}, \beta_{ik})Y_{jk}(\alpha_{jk}, \beta_{jk}) = -Y_{jk}(\alpha_{jk}, \beta_{jk})Y_{ik}(\alpha_{ik}, \beta_{ik})Y_{ij}(\alpha_{ij}, \beta_{ij}).
\end{eqnarray}
It is easy to check that along with $M=A$ or $M=B$, the anti-Yang-Baxter operator $Y$ satisfies the spectral parameter dependent constraint equations
\begin{eqnarray}
     & M_iM_jY_{ij}(\alpha_{ij}, \beta_{ij}) = -Y_{ij}(\alpha_{ij}, \beta_{ij})M_jM_i & \nonumber \\
    & Y_{ij}(\alpha_{ij}, \beta_{ij})Y_{kl}(\alpha_{kl}, \beta_{kl})Y_{ik}(\alpha_{ik}, \beta_{ik})Y_{jl}(\alpha_{jl}, \beta_{jl}) & \nonumber \\
    & =Y_{jl}(\alpha_{jl}, \beta_{jl})Y_{ik}(\alpha_{ik}, \beta_{ik})Y_{kl}(\alpha_{kl}, \beta_{kl})Y_{ij}(\alpha_{ij}, \beta_{ij}).&
\end{eqnarray}
Thus these spectral parameter dependent $Y$ matrices can be lifted to spectral parameter dependent higher simplex equations.

\section{Conclusions}
\label{sec:conclusion}
In this work we have systematically shown how to construct higher simplex operators from lower simplex ones. This involves splitting up an operator with $d$ indices into constituent operators, each with a lower number of indices. Among the many choices for these constituents, we take the simplest ones and demand that they solve the $d$-simplex equation. This naturally leads to constraint equations that resemble the simplex equations. These constraints are shown to be satisfied by operators constructed out of Majorana and Dirac fermions, and those constructed from a pair of anticommuting operators realized using Clifford algebras. The elegance of this method lies in the fact it can provide classes of solutions of the $d$-simplex equation for all $d$.

Here is a quick guide to where the main results are located :
\begin{itemize}
    \item The constraint equations YM-, YY-, TM-, TY- and TT- can be found in \eqref{eq:YMMconstraint-1}, \eqref{eq:YYconstraint}, \eqref{eq:TMMMconstraint-1}, \eqref{eq:TYconstraint} and \eqref{eq:TTconstraint} respectively. While the former two are discussed in Sec. \ref{sec:constraints}, the latter three are written down in Sec. \ref{sec:tetrahedron}. The most general set of constraints appearing while lifting an arbitrary $d$-simplex operator are given by \eqref{eq:jjconstraint} and \eqref{eq:ljconstraint} in Sec. \ref{sec:generalLifts}.
    \item The Majorana fermion 2- and 1-simplex operators that can be lifted are given by \eqref{eq:Ymajorana} and \eqref{eq:Mmajorana} respectively. The Dirac (complex) fermion 3-, 2- and 1- simplex operators that can be lifted are shown in \eqref{eq:TYMfermions}, \eqref{eq:Yfermion-1}, \eqref{eq:Yfermion-2} and \eqref{eq:Mfermion}. In \eqref{eq:TYMclifford} we have a set of Clifford solutions that can be lifted. 
    \item Anti-lower simplex operators can also be lifted. This is discussed in Sec. \ref{sec:lifting-antiYBO}. This leads to more examples of tetrahedron and other higher simplex solutions. Once again we use Majorana fermions, Dirac fermions and Clifford algebras to construct the solutions. Relevant examples can be found in \eqref{eq:anti-Majorana}, \eqref{eq:anti-Majorana-5-simplex} for the Majorana case, \eqref{eq:anti-DiracY} along with the associated $M$ operator for the Dirac case and in \eqref{eq:anti-CliffordY} and their associated $M$'s for the Clifford case.
    \item The higher simplex solutions obtained in this work can be used to construct higher dimensional statistical mechanics models that are integrable. This is especially true for the models constructed using Clifford algebras. This is discussed in Sec. \ref{sec:positivity} where it is shown that three of the solutions from Table \ref{tab:summary}, namely the $H1,4$, $H2,3$ and $H3,1$ classes of solutions give rise to positive Boltzmann weights and thus physically meaningful statistical mechanics models that are integrable.
\end{itemize}

We close with the following comments that also outline the scope for future work.
\begin{enumerate}
    \item The Majorana braid operator ($\sigma_{ij}$) in \eqref{eq:majoranaBraid} has an action on the Majorana operators $\gamma_k$ that can be interpreted as the braiding of Majorana fermions \cite{Ivanov_2001}. The action of $\sigma_{ij}$ is defined by
    \begin{eqnarray}\label{eq:MajoranaAction}
        S_\sigma(\gamma_k) = \sigma_{ij}\gamma_k\sigma_{ij}^{-1}.
    \end{eqnarray}
    With this action we have
    \begin{eqnarray}
        S_\sigma(\gamma_i) = -\gamma_j~;~S_\sigma(\gamma_j) = \gamma_i~;~S_\sigma(\gamma_k) = \gamma_k~~k\neq i,j,
    \end{eqnarray}
     that demonstrates the desired effect of the Majorana braid operators. Along these lines we can compute the action of the tetrahedron and other higher simplex operators on the Majorana fermions. For instance the pure (non-lifted) Majorana tetrahedron operator $T_{ijk}$ in \eqref{eq:Tmajorana} leaves $\gamma_i$, $\gamma_j$ and $\gamma_k$ invariant but on $\gamma_l$ it has a non-trivial action
     \begin{eqnarray}
         S_T(\gamma_l) = \gamma_i\gamma_j\gamma_k\gamma_l,~~l\neq i,j,k.
     \end{eqnarray}
     On the other hand the lifted tetrahedron operator
     \begin{eqnarray}
         T_{ijk} = P_{ij}\left[\mathbb{1}+\gamma_i\gamma_j\right]\gamma_k,
     \end{eqnarray}
     leaves $\gamma_i$ and $\gamma_k$ invariant. They also leave $\gamma_j$ and $\gamma_l$ invariant but with the sign reversed. We can also consider a modified tetrahedron operator
     \begin{eqnarray}\label{eq:braidTmajorana}
         \tilde{T}_{ijk} = \left[\mathbb{1}+\gamma_i\gamma_j\right]\gamma_k,
     \end{eqnarray}
     that satisfies a modified tetrahedron equation 
     \begin{eqnarray}\label{eq:braided-3-simplex}
         \tilde{T}_{kmn}\tilde{T}_{ijn}\tilde{T}_{jlm}\tilde{T}_{ijk} = \tilde{T}_{jlm}\tilde{T}_{ijk}\tilde{T}_{jln}\tilde{T}_{kmn}.
     \end{eqnarray}
     This equation resembles a `braided' version of the vertex form of the 3-simplex equation \eqref{eq:3-simplex}. As we have 
     \begin{eqnarray}
         \tilde{T}_{ijk} = P_{ij}T_{ijk},
     \end{eqnarray}
     the two tetrahedron equations can be seen as analogous to the braided and non-braided forms of the Yang-Baxter equation. The `braid' tetrahedron operator $\tilde{T}_{ijk}$ interchanges $\gamma_i$ and $\gamma_j$, but with the signs reversed when compared to the action of the Majorana braid operator \eqref{eq:majoranaBraid}. It leaves $\gamma_k$ and $\gamma_l$ invariant, the latter with the sign reversed. 

     The tetrahedron operator \eqref{eq:anti-Majorana} obtained by lifting the anti-Yang-Baxter Majorana solution that satisfies the modified tetrahedron equation in \eqref{eq:braided-3-simplex} is 
     \begin{eqnarray}
         \tilde{T}_{ijk} = \left(\gamma_i+\gamma_j\right)\gamma_k~;~\tilde{T}_{ijk}^{-1} = \frac{\left(\gamma_i+\gamma_j\right)}{2}\gamma_k.
     \end{eqnarray}
     Their action on the Majoranas resembles a permutation operator with 
     \begin{eqnarray}
         S_T(\gamma_i)=\gamma_j~;~S_T(\gamma_j)=\gamma_i~;~S_T(\gamma_l)=\gamma_l,~~l\neq i,j.
     \end{eqnarray}
     The tetrahedron operators obtained from the Dirac fermions also have an action on the Majoranas through the expression \eqref{eq:d2m}. These are far more complicated when compared to the above actions. We can also study the action of other higher simplex operators, especially the 4-simplex operator on the Majoranas and check their eventual application to topological quantum computing \cite{cite-key} {\it via} $p$-wave superconductors. These issues will be addressed in a separate paper.
     \item We have seen several examples of spectral parameter dependent Yang-Baxter operators that can be lifted. The resulting three dimensional integrable systems can be interpreted as those having boundary theories described by a two dimensional integrable model governed by the lifted $Y$ operator. More speculatively these methods could help us systematically check the holography principle \cite{Castro2011TheGD}\footnote{We thank Adarsh Sudhakar for pointing out this reference. }.
\end{enumerate}

\section*{Acknowledgments}
PP thanks Somnath Maity, Akash Sinha and Tinu Justin for useful discussions. We also thank the anonymous Referee of Nuclear Physics B in helping us improve the manuscript.

VK is funded by the U.S. Department of Energy, Office of Science, National Quantum Information Science Research Centers, Co-Design Center for Quantum Advantage ($C^2QA$) under Contract No. DE-SC0012704. 



\bibliographystyle{acm}
\normalem
\bibliography{refs}

\end{document}